\documentclass[usenatbib,usegraphicx]{mn2e}

\usepackage{times}
\usepackage{txfonts}
\usepackage{verbatim}
\usepackage{lscape}

\def\aaps{A\&AS}
\def\aap{A\&A}
\def\apj{ApJ}
\def\pasp{PASP}
\def\araa{ARA\&A}
\def\aj{AJ}

\begin{document}

\title{The Na I D resonance lines in main sequence late-type stars}
\author[D\'iaz et al.]{Rodrigo F. D\'\i az, Carolina Cincunegui and Pablo J. D. Mauas \\ Instituto
de Astronom\'\i a y F\'\i sica del Espacio, CC. 67, suc. 28,
1428. Buenos Aires, Argentina.  }
   
 \maketitle

\begin{abstract}
We study the sodium D lines (D1: 5895.92 \AA; D2: 5889.95 \AA) in
late-type dwarf stars. The stars have spectral types between F6 and
M5.5 ($B-V$ between 0.457 and 1.807) and metallicity between $[Fe/H] =
-0.82$ and $0.6$. We obtained medium resolution \emph{echelle} spectra
using the 2.15-m telescope at the argentinian observatory CASLEO. The
observations have been performed periodically since 1999. The spectra
were calibrated in wavelength and in flux. A definition of the
pseudo-continuum level is found for all our observations. We also
define a continuum level for calibration purposes. The equivalent
width of the D lines is computed in detail for all our spectra and
related to the colour index $(B-V)$ of the stars. When possible, we
perform a careful comparison with previous studies. Finally, we
construct a spectral index ($R^\prime_D$) as the ratio between the
flux in the D lines, and the bolometric flux. We find that, once
corrected for the photospheric contribution, this index can be used as
a chromospheric activity indicator in stars with a high level of
activity. Additionally, we find that combining some of our results, we
obtain a method to calibrate in flux stars of unknown colour.
\end{abstract}


\section{Introduction}
All through the main sequence, the Na I D resonance lines (D1: 5895.92
\AA; D2: 5889.95 \AA) are ubiquitous absorption features, clearly
visible in stars of all spectral types. In particular, for cool stars
at the end of the main sequence (late G, K and M) the doublet develops
strong absorption wings. In the most active flare stars, the D lines
show chromospheric emission in their core, a telltale of collision
dominated formation processes.

Besides the intrinsic interest in understanding how these lines behave
for different stellar parameters, the sodium doublet provides a useful
diagnostic tool when studying stellar atmospheres. In this regard,
\citet{andretta97}, \citet{shortdoyle98} and \citet{mdwarves2} showed
that in M dwarfs the sodium D lines provide information of the
conditions in the middle-to-lower chromosphere, and therefore
complements the diagnostics of the upper-chromosphere and low
transition region provided by H$\alpha$. \citet{tripi} modelled the
equivalent width of the doublet in stars of a wide range of spectral
types (between F6 and M5.5). They concluded that the chromosphere is
not very important in determining the equivalent width, since it only
affects emission in the central core, which provides a small
contribution to the doublet strength. However, their model fails to
reproduce the tendency shown by their observations for stars with
$T_{eff}<4000^\circ K$.

In the present work we study different features of the D lines using
medium resolution \emph{echelle} spectra covering the entire visible
spectrum. Our study is focused on stars at the end of the main
sequence (from late F to middle M), for which we analyze the continuum
and line fluxes and the equivalent width of the doublet.

In Sect.~2 we describe the observations and calibration process. In
Sect.~3 we study the continuum flux near the D lines. The equivalent
width is treated in Sect.~4, where we also describe a method to obtain
an approximate flux calibration for stars of unknown colour index. In
Sect.~5 we study the changes observed in the D lines as a consequence
of chromospheric activity. Finally, in Sect.~6 we discuss the results
and present our conclusions.

\section{Observations and stellar sample}

\subsection{Observations and calibration of the spectra}
The observations were obtained at the Complejo Astron\'omico El
Leoncito (CAS\-LEO), in the province of San Juan, Argentina, in 27
observing runs starting in March 1999.  At present, the spectra
of only 19 of these runs were reduced and calibrated.

We used the 2.15-m telescope, equipped with a REOSC \emph{echelle}
spectrograph designed to work between 3500 \AA\ and 7500 \AA. As a
detector we used a 1024 x 1024 pixel TEK CCD.

We obtained spectra covering the whole region between 3860 and 6690
\AA, which lies in 24 \emph{echelle} orders. Due to this complete
coverage we can study the effect of chromospheric activity in all the
optical spectrum simultaneously. In the present work we concentrate on
the sodium D lines. The study of other lines is done somewhere else
\citep[see][]{indices, magnesio}.

We used a 300 $\mu m$-width slit which provided
a resolving power of $R = \lambda/\delta\lambda \approx 26400$. This
corresponds to a spectral resolution of around $0.22$ \AA\ at the
centre of the D doublet. Additionally, we obtained ThAr spectra for
the wavelength ca\-li\-bra\-tion and medium resolution long slit spectra
that were employed for the flux calibration.

The observations were reduced and extracted using standard
    IRAF\footnote{IRAF is distributed by the National Optical
    Astronomy Observatories, which are operated by the Association of
    Universities for Research in Astronomy, Inc., under cooperative
    agreement with the National Science Foundation.}  routines. The
    details of the flux calibration process are presented in
    \citet{library}.

\subsection{Stellar sample}
Since 1999 we have been monitoring around 110 stars of the main
sequence on a regular basis. Each star was observed around three
times a year, weather permitting.

The stellar sample is presented in Table~\ref{todas}. The stars have
  spectral types between F6 and M$5.5$ and colour indexes $(B-V)$
  between $0.457$ and $1.807$, from the Hipparcos/Tycho catalogues
  \citep{hipparcos,tycho}. The metallicity was obtained from different
  sources, as shown in Table~\ref{todas} and range from $[Fe/H] =
  -0.82$ to $0.6$.

The sample was gathered for different research programmes. All the
  stars are field stars, and were included either for being similar to
  the Sun, for exhibiting high levels of chromospheric activity -as is
  usually the case for the M stars-, or for having planetary
  companions. We have also observed a few sub-giant stars for
  calibration purposes, which were excluded from the present
  analysis. In previous works we used 18 of the stars included in the
  sample to calibrate the S index calculated at CASLEO \citep{cycles,
  indices} to the one obtained at Mount Wilson \citep{MW}. These 18
  stars also belong to the group of standard stars used by
  \citet{CTIO} at Cerro Tololo. Several stars were included to cover
  the entire range of effective temperature. However, the sample is
  not uniform in metallicity or effective temperature. Nevertheless,
  we believe our results are not biassed by this lack of uniformity in
  our sample. When such a risk is present, we binned the observations
  (see Sect.~\ref{pc}).

Our sample also includes seven stars that fall outside the main
sequence when placed in an HR diagram, even though they are classified
as dwarf stars in the SIMBAD\footnote{http://simbad.u-strasbg.fr}
database. For these stars, we calculated the absolute magnitude in the
V filter ($M_V$) using the measurements of parallax and V from the
Hipparcos/Tycho catalogues \citep{hipparcos,tycho}. We list these
stars in Table~\ref{tabla1}. The fourth column shows the luminosity
class that corresponds to the calculated $M_V$
\citep[see][$\S95$]{allenviejo}, obtained assuming that the spectral
type is correct. These stars were excluded from the analysis, together
with the rest of the non-main-sequence stars.

Also contained in our sample are four stars which are listed in a
 luminosity class other than V in the SIMBAD database that actually
 fall well whithin the main sequence. Again, $M_V$ values were
 obtained and compared with the table in \citet{allenviejo} to confirm
 their placement in the HR diagram. These stars are shown in
 Table~\ref{tabla2}, and were included in the analysis. We also
 excluded from the analysis the stars with unknown parallax, since for
 them absolute fluxes cannot be determined. Our final sample consists
 of 652 spectra for 84 stars.
\renewcommand{\arraystretch}{1.2}
\begin{table}
\begin{center}
\begin{tabular}{l c c c c}
\hline
&Spectral type&$(B-V)$&$M_V$&Lum Class\\
\hline \hline
hd103112&K0&1.05&3.18&IV \\
hd105115&K2/K3&1.41&-0.68&III\\
hd119285&K1p&1.04&3.39&IV\\
hd25069&G9&0.99&2.75&IV \\
hd5869&K4&1.49&1.05&III\\
hd94683&K4&1.78&-2.38&II\\
hd17576&G0&1.78&1.91&III\\
\hline
\end{tabular}
\caption{\label{tabla1} Stars located outside the main sequence in an
  HR diagram despite being classified as dwarf-type stars. The fourth
  column is the luminosity class we obtained assuming the spectral
  type is correct}
\end{center}
\end{table}
\begin{table}
\begin{center}
\begin{tabular}{l c c c}
\hline
Star&Spectral type (SIMBAD)&$(B-V)$&$M_V$\\
\hline \hline
hd52265&G0 III-IV&0.57&4.12\\
hd57555&GO IV/V&0.66&4.18\\
hd75289&G0 I&0.58&4.1\\
hd120136&F6 IV&0.51&3.58 \\
\hline
\end{tabular}
\caption{\label{tabla2} Stars located in the main sequence
  despite not being classified as dwarf-type stars.}
\end{center}
\end{table}

\section{Pseudo-continuum}
\label{pc}
\subsection{Definition \label{contdef}}
\begin{figure*}
         \includegraphics[width=\textwidth]{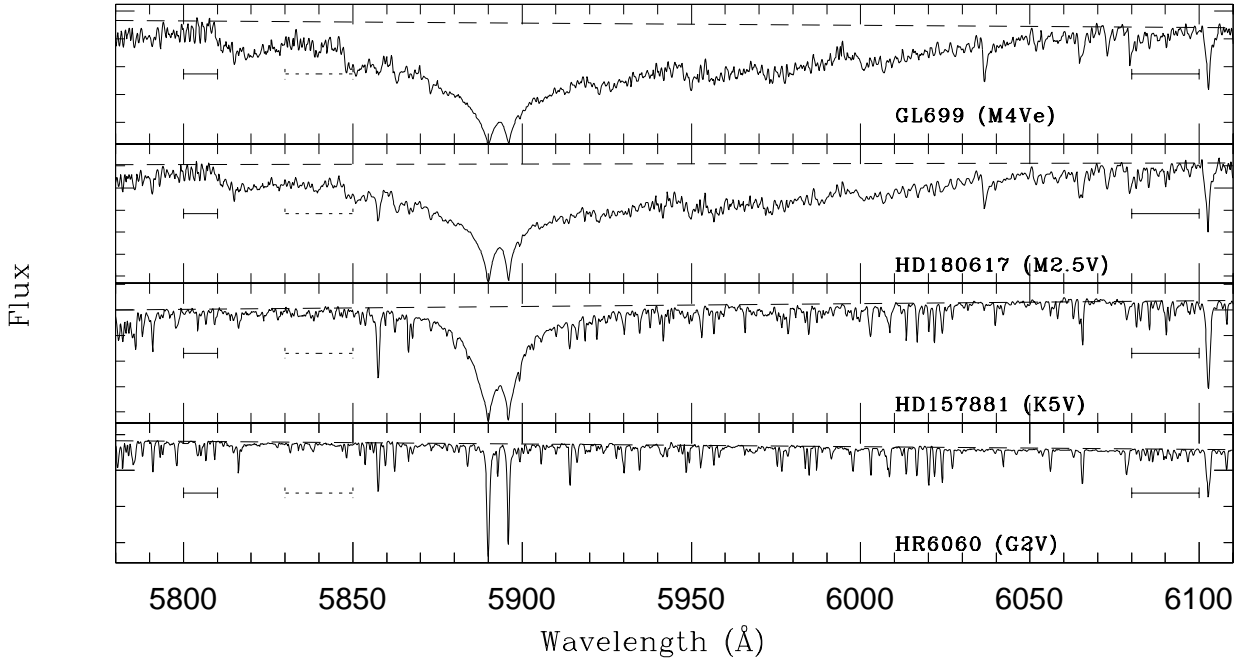}
         \caption{The region of the sodium doublet for stars of different
         spectral types. The dotted lines mark the windows used for the
         calculation of the continuum flux. The dashed line is the continuum
         level as defined in Sect.~\ref{contdef}.}
         \label{ufig1}
         \end{figure*}

In Fig.~\ref{ufig1} we show the spectrum in the region of the D
lines for four stars of different spectral types. As can be seen in
the spectra, in the case of later stars the doublet forms inside a
strong molecular band, which extends from 5847 \AA\ to 6058 \AA\ and
is due to TiO~\citep{adleo1}

Therefore, in order to obtain a reliable estimate of the continuum
flux near the lines for all spectral types, the windows chosen have to
be located far away from the line centre, outside the molecular
bands. The closest regions are a 10 \AA\ wide window centered at 5805
\AA\ (window V) and a 20 \AA\ wide window centered at 6090 \AA\
(window R). These windows are indicated in Fig.~\ref{ufig1} with solid
lines.

However, even in these regions there are several photospheric lines,
and therefore computing the mean flux would lead to a considerable
underestimation of the continuum flux. To deal with this problem, we
considered the mean value of the ten highest points in each window as
an indication of the continuum flux in it. Finally, to take into
account the shape of the continuum in this region, we computed the
flux at the doublet centre ($F_{cont}$) as the linear interpolation of
the values in each window.

In Fig.~\ref{ufig1}, the linear interpolation in each spectrum is shown
as a dashed line. We see that this is a good estimation for the continuum flux
at the centre of the doublet, even for M-type stars.

\subsection{Relation with (B-V)}

\begin{figure}
        \centering
        \includegraphics[width=\columnwidth]{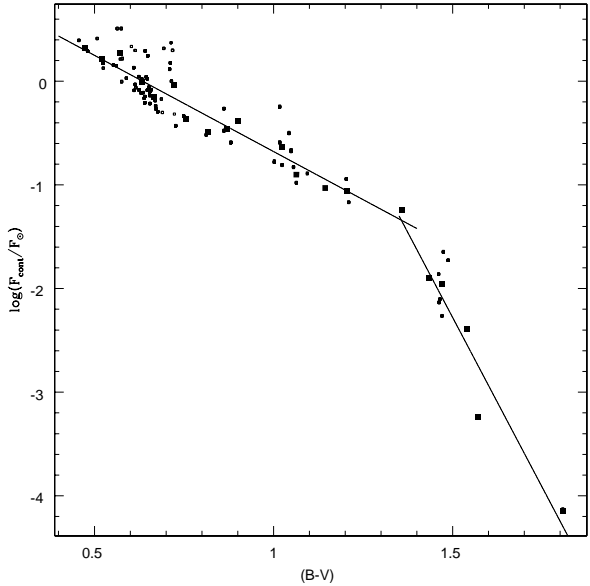}
        \caption{$\log(F_{cont}/F_\odot)$ vs. colour index
         $(B-V)$. The small empty squares represent the individual
         observations, and the filled squares are the mean value in
         each bin (see text). The solid lines are the best piecewise
         fit to the data. Note that in several bins there is only
         one star, and the individual observations are therefore
         hidden by the filled square.}
        \label{ufig7}
        \end{figure}

We calculated $F_{cont}$ for all the spectra in our sample in the way
explained above. $F_{cont}$ was normalized to the flux $F_\odot$,
defined in such a way that $F_{cont}/F_\odot = 1$ for a star of solar
colour $(B-V)=0.62$. In Fig.~\ref{ufig7} we plot $F_{cont}/F_\odot$ in
logarithmic scale as a function of colour index $(B-V)$. The trend is
similar to that shown by the lower end of the main sequence in an HR
diagram.

As can be seen in Fig.~\ref{ufig7}, the stars are not evenly
distributed with colour, a fact that might bias our results. To avoid
this, we binned the observations every 0.05 units in $(B-V)$. For each
bin, we computed the mean value of $F_{cont}$, which are plotted as
filled squares in Fig.~\ref{ufig7}. It can be seen in the figure
that there exists a break at $(B-V)=1.4$, where M stars
begin. Therefore, we fitted the binned data with two linear
relations:
\begin{equation}
       log\left(\frac{F_{cont}}{F_{\odot}}\right)= \left\{\begin{array}{c r}
       1.18 - 1.86\,\times\,(B-V)&\ \ \ \ \;(B-V<1.4)\\ \\ 7.55 -
       6.55\,\times\,(B-V)&\ \ \ \ \;(B-V\geq1.4)
       \end{array}\right.
       \label{cal1}
\end{equation}
where $F_\odot = 5.7334\times10^{-11}$
erg\,cm$^{-2}$\,s$^{-1}$\,\AA$^{-1}$. 

\subsection{Calibration}

A relation like that of Eq.~\ref{cal1} can be used to calibrate in
flux spectra of stars of known colour index $(B-V)$. However, most
spectrographs only observe a small wavelength range around the line of
interest. For these instruments, the windows at 5805 \AA\ and 6010
\AA\ might fall outside the observed range. For this reason, we
studied a relation similar to the one in Eq.~\ref{cal1} for a
continuum window closer to the doublet.

First, we note that due to the presence of molecular bands, the raw
spectra of M-type stars are difficult to normalize, in particular if
the observed range is not wide enough. Therefore, we did not try
to find a good pseudo-continuum to calibrate M-dwarfs. For F, G and K
stars, we considered a window 20 \AA\ wide around 5840 \AA, which is
indicated in Fig.~\ref{ufig1} with dotted lines.

\begin{figure}
        \centering
        \includegraphics[width=\columnwidth]{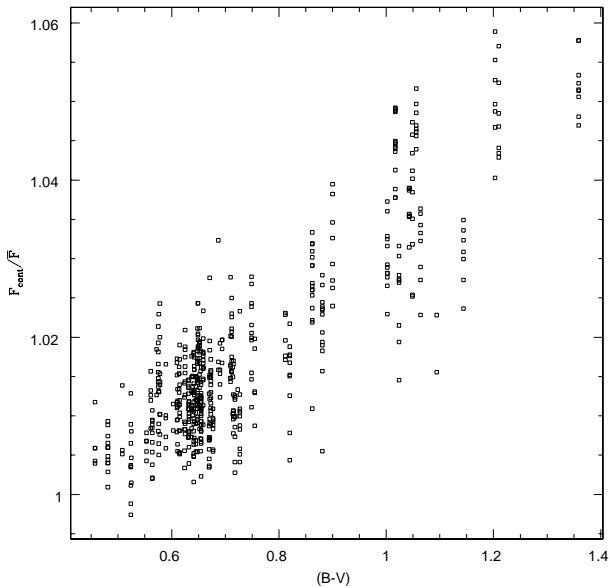}
        \caption{Ratio between $F_{cont}$ and the mean flux $\bar{F}$ in the
          5830-5850 \AA\ window, as a function of colour index $(B-V)$. The
          percentage difference remains smaller than 6 per cent for F, G and K
          stars.}
        \label{coefcont}
        \end{figure}

In Fig.~\ref{coefcont} we show the ratio between the mean flux
$\bar{F}$ in this window and the one obtained earlier, for stars with
$(B-V)<1.4$. As can be seen, using $\bar{F}$ underestimates the true
continuum, but the percentual difference between both is
smaller than 6 per cent. Therefore, the first part of Eq.~\ref{cal1} can be
used to fit the mean flux $\bar{F}$, and to calibrate in flux this
region of the spectrum of a star of known $(B-V)$, increasing the
error by less than 6 per cent.

\section{Equivalent width}
\label{sec.an}

\subsection{Definition \label{definition}}

The equivalent width $W_\lambda$ of a spectral line is defined as
\begin{equation}
        W_\lambda = \int_{\lambda_1}^{\lambda_2}
        \left(1 - \frac{F_\lambda}{F_c}\right) \,d\lambda \ \ ,
        \label{ancho}
        \end{equation}
where $F_c$ is the continuum flux and $F_\lambda$ is the value of the
flux in the line. As we pointed out in the preceeding section, the choice
of the continuum level can be subjective, and it can be very difficult 
for M-dwarfs. 

On the other hand, the limits of the wings of a line are not always
well-defined, and therefore the integration limits $\lambda_1$ and
$\lambda_2$ also constitute subjective points in the definition of
$W_\lambda$. In fact, while in earlier F and G stars the photospheric
wings are thin and well defined, in later stars the wings are
extremely wide. Therefore, the choice of a proper integration window
requires a varying width ($\Delta\lambda = \lambda_2 - \lambda_1$),
ranging from 14 to 40 \AA\ around the centre of the doublet (5892.94
\AA), depending on the spectral type of the star. In
Table~\ref{waveint} we list the values of $\Delta\lambda$ we used for
different colours. A similar technique, with varying widths, was used
by \citet{tripi}, although it is not clear from their paper which
$\Delta\lambda$ was used for each colour.

Note that in stars with spectral type later than M2.5 the absorption
wings of the D lines become very large and blended with the
highly-developed absorption bands present (see
Fig.~\ref{ufig1}). Therefore, for these stars a suitable choice of
$\Delta\lambda$ is very difficult and the calculation of the
equivalent width cannot be considered entirely reliable.

\begin{figure}
        \centering
        \includegraphics[width=\columnwidth]{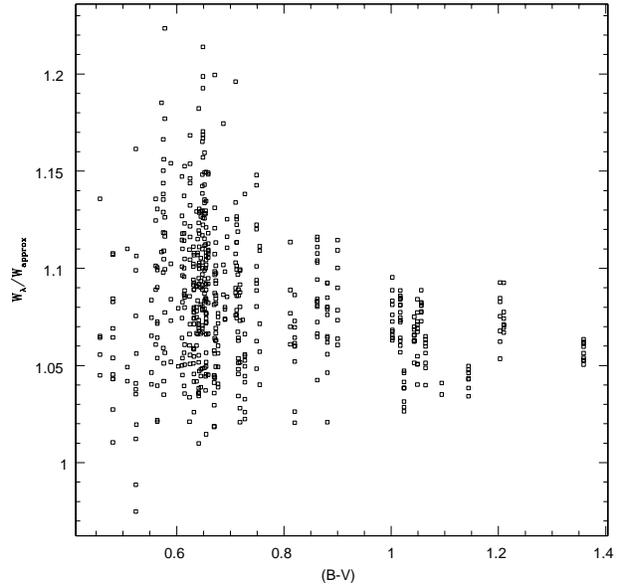}
        \caption{Ratio between $W_\lambda$, calculated with a realistic
          estimation of the continuum flux ($F_c = F_{cont}$), and $W_{approx}$,
          computed using the definition of $F_c$ suitable for the calibration
          process described in Sect.~\ref{calibra} ($F_c = \bar{F}$). The
          percentage difference remains smaller than 20 per cent for almost all stars.}
        \label{coefancho}
        \end{figure}

Regarding the continuum level, we used both values described in the
previous section.  The equivalent width computed with $\bar{F}$, the
aproximate definition of $F_{c}$, will be referred to as $W_{approx}$,
and the one computed with $F_{cont}$ will be refered simply as
$W_\lambda$.  The ratio between both values is plotted in
Fig.~\ref{coefancho}. Again, using $\bar{F}$ leads to an
underestimation of the equivalent width. The differences are larger
than for the continuum flux, but even in this case they remain smaller
than 20 per cent.

\renewcommand{\arraystretch}{1.4}
        \begin{table}
        \begin{center}
        \begin{tabular}{c c}
        \hline
        $(B-V)$ interval&$\Delta\lambda$~[\AA]\\
        \hline \hline
        $\leq0.8$&14\\
        $[0.8-1)$&16\\
        $[1.0-1.2)$&20\\
        $[1.2-1.3)$&24\\
        $[1.3-1.5)$&28\\
        $>1.5$&40\\
        \hline
        \end{tabular}
        \caption{\label{waveint} Wavelength window $\Delta\lambda$ used in
          the calculation of the equivalent width for different values of
$(B-V)$
          (see Equation~\ref{ancho}).}
        \end{center}
        \end{table}

\begin{figure}
        \centering
        \includegraphics[width=\columnwidth]{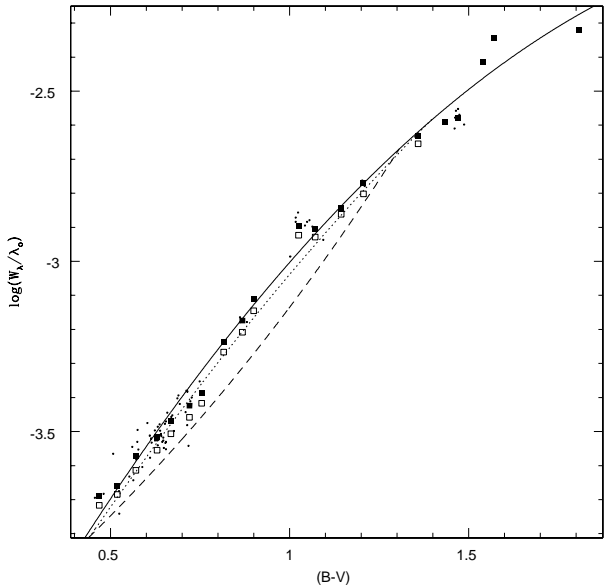}
        \caption{Equivalent width (logarithmic scale) vs $(B-V)$. The
          small dots are the mean values for each star and the filled
          squares are the mean values in each bin. The empty squares
          are the mean values of $W_{approx}$ in each bin (see
          text). The solid line is the best fit to the binned
          data. The dotted line is the fit to $W_{approx}$ and the
          dashed line is the fit given by Tripicchio et al.,
          transformed to the colour index scale using
          Eq~\ref{ecnoyes}.}
        \label{anchocolor}
        \end{figure}

\subsection{Relation with colour index $(B-V)$}

In Fig.~\ref{anchocolor} we illustrate the behaviour of $W_\lambda$ as
a function of the colour index $(B-V)$. In the \emph{y}-axis we plot
$\log(W_\lambda/\lambda_o)$, where $\lambda_o = 5890$ \AA. The small
dots represent the mean value per star and the filled squares
show the results of binning the data as explained in the previous
section. As empty squares we plot the values of $W_{approx}$, binned
in the same way.

A quadratic fit to the binned data gives
\begin{equation}
  \log\left(\frac{W_\lambda}{\lambda_o}\right)= -4.579 +
1.948\,(B-V)-0.3726\,(B-V)^2\ \ .
     \label{anchoa}
     \end{equation}
This fit is shown as a solid line in Fig.~\ref{anchocolor}. It can be
seen that, as the temperature decreases, the D lines become wider. It
is evident from Fig.~\ref{anchocolor} that there is a saturation
effect for the reddest stars. As shown in Fig.~\ref{ufig1}, for M
stars with $(B-V)\geq1.4$, the lines are so wide that they are almost
completely blended. At this point, the doublet behaves as a single
line, and the equivalent width grows mainly in the outer
wings. Additionally, as cooler stars are also more active, the central
line cores are filled in by chromospheric emission, contributing to
the saturation effect.

\subsection{Comparison}
\citet{tripi} also studied the dependence of the equivalent width with
effective temperature. They constructed synthetic profiles of the D
lines using a modified version of the MULTI code, which takes into
account the line blending present in the doublet. They also
observed a sample of dwarf and giant stars with spectral type between
F6 and M5. They used both the modelled atmospheres and the observed
spectra to obtain a relation between $W_\lambda$ and effective
temperature, given by:
\begin{equation}
\log\left(\frac{W_\lambda}{\lambda}\right)=2.43\cdot\frac{5040}{T_{eff}}-5.73\
    \ \ ,
    \label{modelotripi}
    \end{equation}
which reproduces their observations fairly well for effective temperatures
higher than 4000~$^\circ$K. 

In order to compare this relation to our regression we performed a
change of variables in equation~\ref{modelotripi} using the relation
given by \citet{noyes84}:
\begin{equation}
     \log(T_{eff})=\,3.908 - 0.234\cdot(B-V)\,\, ,
     \label{ecnoyes}
     \end{equation}
valid for stars with $0.4\!<(B-V)<1.4$. In this way we obtain
$W_\lambda$ as a function of $(B-V)$. The resulting expression is
plotted as a dashed line in Fig.~\ref{anchocolor}.

It can be seen that the values by \citet{tripi} are smaller than ours:
in some cases the difference between both expressions is as large as
40 per cent. Unfortunately, they do not mention how the continuum level or
the integration windows were chosen. Although they say they used a
wavelength interval of variable width and that they cleared the
spectrum of unwanted lines, no details are given, so we could not
apply their calculation method to our data.

To check whether our larger values of $W_\lambda$ are due to the weak
photospheric lines blended in the profiles, we recomputed the
equivalent width in a completely different way, for several of our
stars. We considered the raw spectra of 16 stars from our sample. The
spectroscopic order in which the D lines are found was extracted,
calibrated in wavelegth and normalized. It is worth noting that the
doublet is located in the middle of this order, where the effects of
the blaze function are less severe. However, the presence of molecular
bands prevented us from normalizing the spectra of M-type stars.

We then computed the equivalent width using the IRAF line-deblending
routines included in the SPLOT task, which fit Voigt profiles to the
lines, and perform the calculation over the fits. In this way, the
presence of photospheric lines has no influence on the obtained
value. Due to the severe blending of the D lines, the routine used by
IRAF did not produce acceptable results for stars with colour larger
than $\approx 1.3$. Therefore, this method can only be applied to
stars with $(B-V) < 1.3$.

\begin{figure}
\centering
\includegraphics[width=\columnwidth]{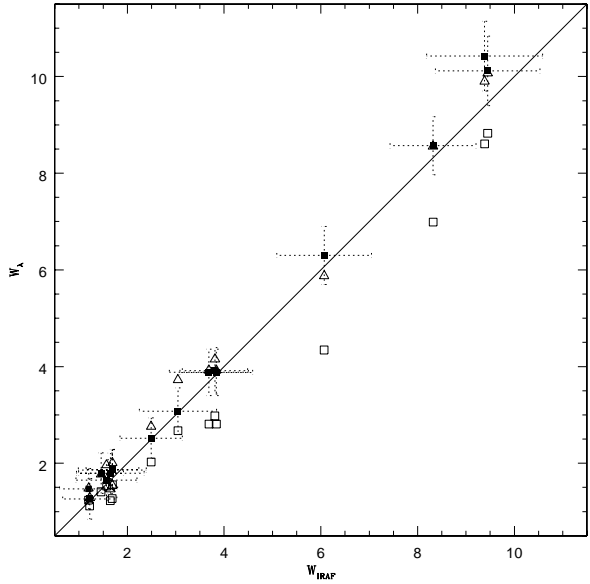}
\caption{$W_\lambda$ computed from the calibrated spectra (filled
  squares) as a funtion of the one given by the IRAF deblend routine
  ($W_\mathrm{IRAF}$). The triangles represent the values obtained
  from the fit of Eq.~\ref{anchoa} and the empty squares represent the
  values obtained using the fit from \citet{tripi}.}
\label{anvsan}
\end{figure}

In Fig.~\ref{anvsan} we plot as filled squares the value of
$W_\lambda$ calculated from our calibrated spectra Vs. the value
obtained using IRAF ($W_{\mathrm{IRAF}}$). The errors in $W_\lambda$
were calculated assuming a 3 per cent error in $F_\lambda/F_c$ (see
Eq.~\ref{ancho}). The error bars in the \emph{x}-axis were taken as
the RMS of the fit done by IRAF times the width of the integration
window (see Table~\ref{waveint}). The solid line represents the
identity relation, and not a fit to the data. It can be seen that
the differences beween our calculation and the one done with IRAF
are well within the errors.  The filled triangles represent the values
of $W_\lambda$ obtained from our fit, rather than from the individual
obervations. Note that the points still remain very close to the
identity line.

As open squares we plot the equivalent width obtained from
Trippichio's relation (Eq.~\ref{modelotripi}) for the value of $(B-V)$
of the stars considered, which in all cases fall below the identity
relation.

\begin{figure}
\centering
\includegraphics[width=\columnwidth]{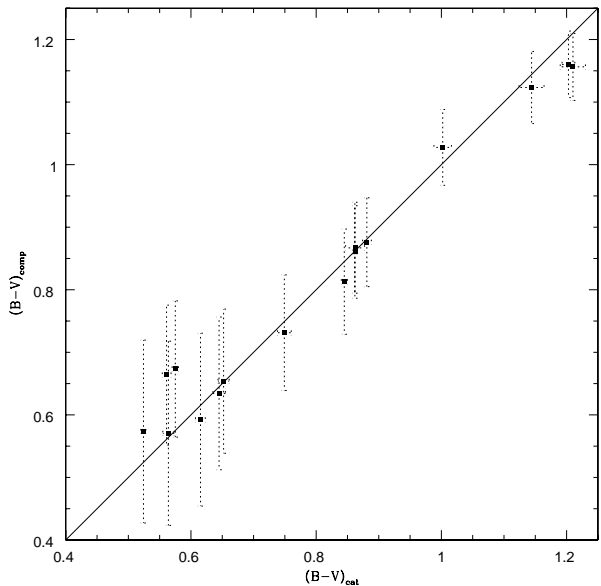}
\caption{$(B-V)$ computed from the normalized ([comp]) spectra through
  the inverse of Eq.~\ref{anchob} as a funtion of $(B-V)$ from the
  Hipparchos/Tycho catalogue ([cat]). The solid line is the identity
  function. A good agreement is found between both values.}
\label{calcolor}
\end{figure}
\subsection{Determination of $(B-V)$\label{calibra}}
The relation shown in Fig.~\ref{anchocolor} can be used to determine
the $(B-V)$ of a star using the equivalent width of the D lines,
measured in non-calibrated spectra. Since in many cases the spectral
region needed to compute $F_{cont}$ might fall outside the observed
one, we repeat the fit for $W_{approx}$. The best least square fit to the
binned values of $\log(W_{approx}/\lambda_o)$ Vs. (B-V) is:
\begin{equation}
    \log\left(\frac{W_{approx}}{\lambda_o}\right)= -4.528 + 1.734\,(B-V) -
0.2442\,(B-V)^2\ \ ,
     \label{anchob}
     \end{equation}
which is valid only for $(B-V)<1.4$, and where $\lambda_o = 5890$ \AA. 
This fit is plotted as a dotted line in Fig.~\ref{anchocolor}. 

To obtain $(B-V)$ from the measured equivalent width, one could invert
Eq.~\ref{anchob}. However, since the least square estimation of
parameters does not treat the dependent and independent variables
symmetrically, a more rigorous procedure would be to fit $(B-V)$ as a
function of $W_{approx}$. We performed this fit and found that within
the errors, both procedures give the same values of $(B-V)$.

The method was verified using the same 16 stars used in the previous
section. The raw spectra were extracted, calibrated in wavelength and
normalized using Chebyshev polynomials. All these procedures were
carried out using IRAF.  $W_\lambda$ was calculated by direct
integration of the flux in the appropiate window (see
Table~\ref{waveint}) using SPLOT. Then, the inverse of
Eq.~\ref{anchob} was used to obtain $(B-V)$ for these stars.

In Fig.~\ref{calcolor} we plot the value of $(B-V)$ computed in this
way ($(B-V)_{comp}$) as a funtion of the value obtained from the
Hipparcos/Tycho catalogue ($(B-V)_{cat}$).  To estimate the error in
$(B-V)_{comp}$ we assumed a 5 per cent error in $F_\lambda/F_c$, in order to
consider errors in the normalization procedure. The errors in the
\emph{x}-axis were taken from the Hipparcos/Tycho catalogue. The solid
line represents the identity relation.  It can be seen that the method
gives, in fact, very good estimates of the colour of the stars.

Therefore, by measuring the equivalent width of the D doublet an estimate
of the colour index $(B-V)$ can be obtained. Afterwards, this value of
$(B-V)$ can be used to obtain an approximate flux calibration in the
region of the D lines, by means of Eq.~\ref{cal1}. In this way, the
spectra of stars of unknown colour can be calibrated in flux in this
spectral region.

\section{Activity indexes and chromospheric activity}
\label{sec.activ}
In this section, we define different activity indexes using the D
lines, and study their applicability as chromospheric activity
indicators.

\subsection{N index \label{nindex}} 
First, we constructed an index (N) similar to the Mount Wilson S
index, described by \citet{MW}.
\begin{displaymath}
N=\frac{f_{1} +
   f_{2}}{f_{cont}}\ \ ,
\end{displaymath}
where $f_i$ is the mean flux in the D$_i$ line, integrated in a square
window 1 \AA\ wide, and $f_{cont}$ is the mean flux of the continuum
calculated as explained in Sect.~\ref{contdef}.

\begin{figure}
\centering
\includegraphics[width=\columnwidth]{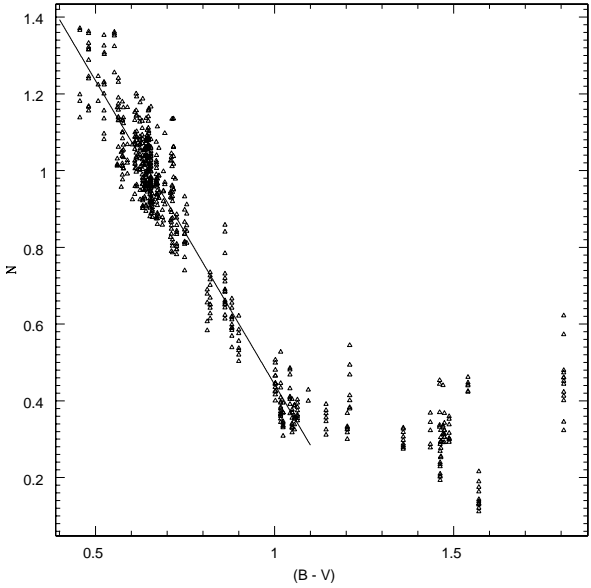}
\caption{N index vs. colour index $(B-V)$. The solid line is
  the best linear fit for stars with $(B-V)<1.1$. Stars with a larger
  $(B-V)$ do not show any correlation.}
\label{ufig3}
\end{figure}

For the earlier stars in our sample, the N index is highly dependent
on the colour $(B-V)$, as can be seen in Fig.~\ref{ufig3}. This
behaviour changes for stars with $(B-V)\ga 1.1$, for which the value
of N is roughly constant. Stars with $(B-V)<1.1$ were fitted with a
linear relation, shown in the figure as a solid line:
\begin{equation}
N_{(B-V)}=-1.584\cdot(B-V)+ 2.027\ .
\label{ec2}
\end{equation}

\begin{figure}
\centering
\includegraphics[width=\columnwidth]{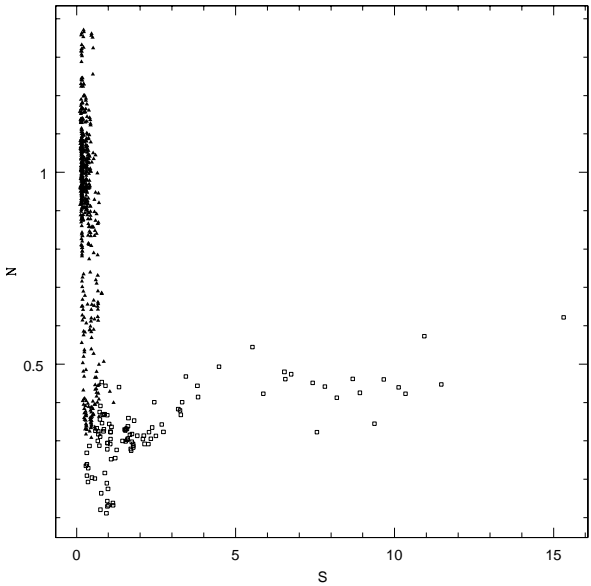}
\caption{N vs. S. Values for stars with $(B-V)<1.1$ are shown as filled
  triangles, and those with $(B-V)\geq1.1$ as empty squares.}
\label{ufig4}
\end{figure}

In Fig.~\ref{ufig4} we show the result of plotting N as a function of
the S index, which was obtained from our spectra using the calibration
presented in \citet{indices}, which relates the S index obtained at
CASLEO with the one computed at Mount Wilson. Note that since our
spectra cover the entire visible range, we measure N and S
simultaneously.

\begin{figure}
\centering
\includegraphics[width=\columnwidth]{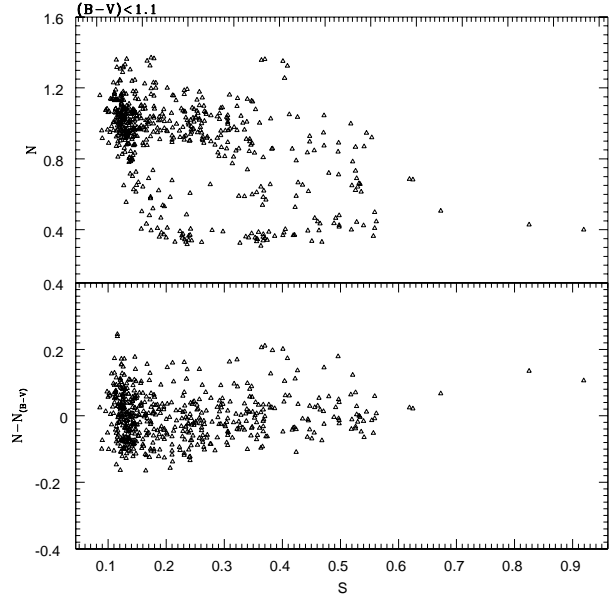}
\caption{\emph{Upper panel}: N vs. S for stars with
  $(B-V)<1.1$. \emph{Lower panel}: The dependence on N with colour has
  been corrected by substracting N$_{(B-V)}$, shown in
  Eq.~\ref{ec2}. The correlation disappears, as is confirmed by the
  small value of the correlation coefficient ($r=0.07$).}
\label{cor_desap}
\end{figure}
\begin{figure}
\centering
\includegraphics[width=\columnwidth]{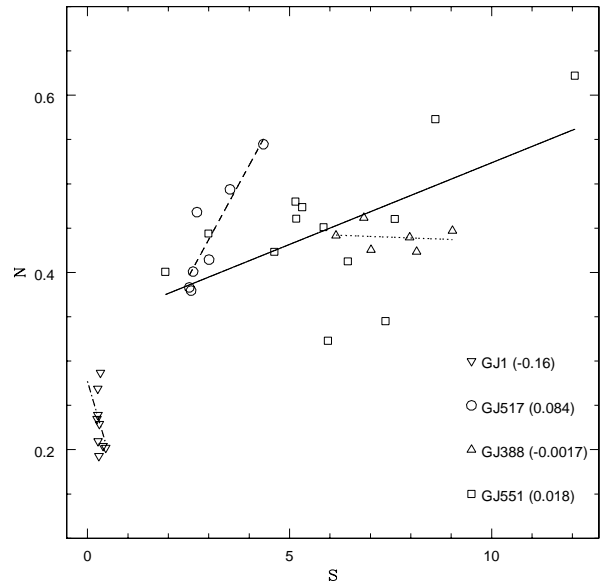}
\caption{N vs S for selected stars with $(B-V)\geq1.1$. A linear fit
  to the data of each star is also shown (GJ551: solid line; GJ388:
  dotted line; GJ517: dashed line; GJ1: dot-dash line). The slope of the fit is
  given in parentheses in the legend.}
\label{nstar}
\end{figure}

It can be seen that stars with $(B-V)<1.1$ (shown as filled triangles)
have smaller values of S, and show a weak anticorrelation between N
and S ($r=-0.488$). We plot the observations for these stars alone in 
the upper panel of Fig.~\ref{cor_desap}. Despite the considerable scatter,
the anticorrelation is evident. 

However, for these stars the correlation between N and $(B-V)$ is
stronger.  To correct for this dependence we substracted
$\mathrm{N}_{(B-V)}$ (Eq.~\ref{ec2}) from N. In the lower panel of
Fig.~\ref{cor_desap} we plot $\mathrm{N}-\mathrm{N}_{(B-V)}$
vs. (B-V). In this case, the correlation dissapears ($r=0.07$),
implying that the correlation was due only to the dependence of N with
colour, and the tendency of cooler stars to be more active.  Threfore,
the N index cannot be used as an activity indicator for stars with
$(B-V)<1.1$.

On the other hand, stars with $(B-V)\geq1.1$ present a good
correlation between both N and S, as can be seen in
Fig.~\ref{ufig4}. Since for these stars N do not correlate with colour,
we do not perform any correction in this case. 

Since we have simultaneous observations of both indexes, we can
explore this apparent correlation for particular stars. In
Fig.~\ref{nstar} we plot N vs S for several stars with $(B-V)\geq1.1$,
and we show the linear fits to each individual dataset (the slope of
each fit is given in the legend).  It can be seen that the slope of
the relation between S and N changes from star to star, and the
behaviour ranges from correlations with different slopes (GJ551 and
GJ517) to cases where N and S are uncorrelated, either with
constant S and variable N (GJ1) or viceversa (GJ388).  A similar
effect was observed comparing the flux of the H$_\alpha$ line to the
flux in the H and K lines \citep{indices}. In that case, we found
correlations with different slopes, anti-correlations, and cases where
no correlation is present.

Therefore, the N index cannot be used to compare activity levels of
different stars. However, it can be useful to compare activity levels
on the \emph{same} star at different times, for those stars which show
a correlation between S and N. For example, N can be used to trace
activity cycles in late-type stars, as we did with Proxima Centauri
using the flux of the H$_\alpha$ emission \citep[see][]{proxima}.

\subsection{$R^\prime_D$ \label{rds}}

\citeauthor{noyes84}
(1984, hereafter N84)\defcitealias{noyes84}{N84}noted that the S
index is sensitive to the flux in the continuum windows and to the
photospheric radiation present in the H and K bandpasses, both of
which depend on spectral type. Despite the apparent independence of 
N with $(B-V)$ seen in Fig.~\ref{ufig3} for stars with $(B-V)\geq1.1$, 
the results of the previous section seem to indicate that N also shows a
dependence on photospheric flux, even for stars with $(B-V)\geq
1.1$. This dependence must be corrected more thoroughly in order to
compare stars of different spectral type.

Following \citetalias{noyes84}, we define a new index 
\begin{equation}
R^\prime_{D} = R_D - R_D^{phot}\; \; ,
\end{equation}
where $R_D = (F_{1}+F_{2})/\sigma\,T_{eff}^4$, $F_{1}$ and $F_{2}$ are the
fluxes in the D1 and D2
windows at the stellar surface, and $\sigma$ is the
Stefan-Boltzmann constant.  Note that $R_{D}$ can be expressed as:
\begin{equation}
R_D=\frac{f_{1}+f_{2}}{f_{bol}}=\frac{f_{cont}}{f_{bol}}\cdot N\; \; ,
\label{rd}
\end{equation}
where $f_{bol}$ is the relative bolometric flux. Both expressions in
Eq.~\ref{rd} have the virtue of depending on fluxes on Earth instead
of fluxes at the stellar surface. $f_{bol}$ was computed interpolating
the values of the bolometric correction provided
by~\citet{johnson66}. Defined in this way, $R_D$ is independent of the
photospheric flux in the continuum window.

$R_D^{phot}$ is the photospheric contribution to the flux in the windows 
of the D lines. For the H and K lines, \citetalias{noyes84} integrated 
the flux outside the H$_1$ and K$_1$ minima, and used these values to
calculate $R_{phot}$, which is substracted from $R_{HK}$. However,
the D lines are usually in absorption. For this reason, we had to correct for
the photospheric contribution in a different way. 
Since even for very inactive stars the D lines are not
completely dark, the minimum line flux in a very inactive star must
be photospheric in origin. Therefore, we assumed that the basal flux
of the D lines is a reasonable estimation of the photospheric
contribution.

\begin{figure}
\centering
\includegraphics[width=\columnwidth]{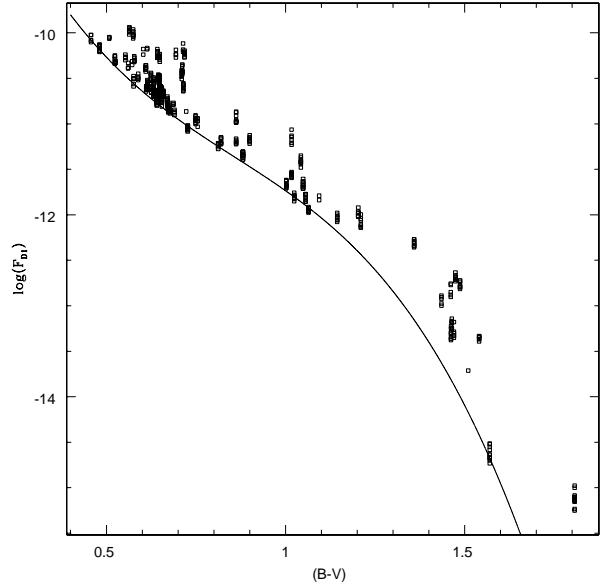}
\caption{Flux in the D1 line as a function of colour index
  $(B-V)$. The solid line represents our estimation of the minimum
  flux. Those spectra with D lines in emission were not considered for
  the fit.}
\label{dflux}
\end{figure}

In Fig.~\ref{dflux} we plot the flux in the D1 line as a function of
the colour index $(B-V)$, in logarithmic scale. The solid line represents
the basal flux, computed using a third order polynomial. The spectra
with the D lines in emission were excluded from the fit. The
resulting expression is:
\begin{equation}
\begin{array} {c c}
\log(F_1^{min})=-6.527 -11.86\cdot&(B-V) +10.87\cdot(B-V)^2 
\\-4.218\cdot(B-V)^3\; \; .
\end{array}
\label{dmin}
\end{equation}

For the D2 line, a similar expression is obtained:
\begin{equation}
\begin{array} {c c}
\log(F_2^{min})=-6.374 -12.48\cdot&(B-V) +11.37\cdot(B-V)^2
\\-4.344\cdot(B-V)^3\; \; .
\end{array}
\label{dmin2}
\end{equation}

Contrary to what happens for the H and K lines, this term is
appreciable for all the range of our observations. Even for stars with
Balmer lines in emission, the photospheric correction can be up to 50
per cent of the value of $R_D$.

Following \citetalias{noyes84}, we substracted the same value of
$R_D^{phot}$ to all stars of the same spectral type. In this way, we
have constructed an index eliminating the dependence on spectral type
which arises from the photospheric flux in the integration
bandpasses. $R^\prime_D$ should therefore depend exclusively on
chromospheric flux and be independent of spectral type.

\subsection{Chromospheric activity}
According to \citetalias{noyes84}, $R^\prime_{HK} = R_{HK} - R_{phot}$ is proportional
to the fraction of nonradiative energy flux in the convective
zone. Since this flux then heats the chromosphere by means of the
magnetic field, $R^\prime_{HK}$ should be a good activity
indicator.

We computed $R^\prime_{HK}$ for all our spectra using a relation
analogue to Eq.~\ref{rd}. As $R_{phot}$ we used the expression given
in \citetalias{noyes84}, valid in the range $0.44<(B-V)<0.82$,
\begin{equation}
\log(R_{phot})=-4.02-1.4\ (B-V)\; \; .
\label{rphot}
\end{equation}
 As noted by \citetalias{noyes84} the photospheric correction is unimportant for active
 stars. Indeed, we found that the correction is less than 10 per cent
 virtually for all stars with $B-V>1.1$. Therefore, we used Eq.~\ref{rphot}
 for all the range of our observations.
\begin{figure}
\flushleft
\includegraphics[width=\columnwidth]{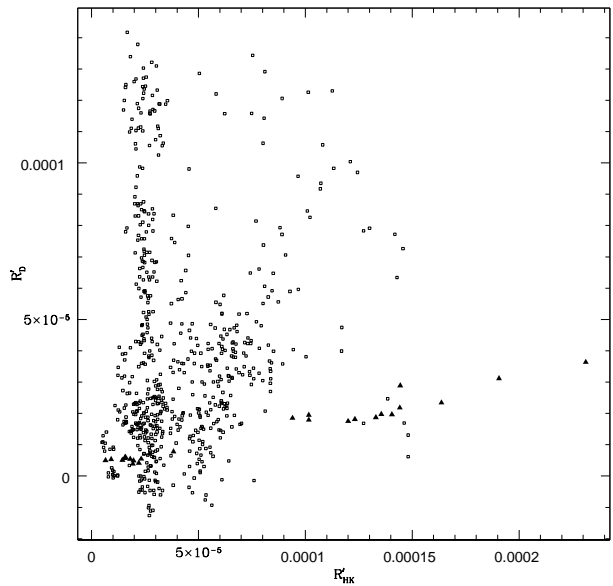}
\caption{$R^\prime_D$ vs $R^\prime_{HK}$. Filled triangles represent
  stars with Balmer lines in emission. No clear correlation is present,
  and the correlation coefficient for the complete sample is $r=0.094$.}
\label{rvsr}
\end{figure}

\begin{figure}
\flushleft
\includegraphics[width=\columnwidth]{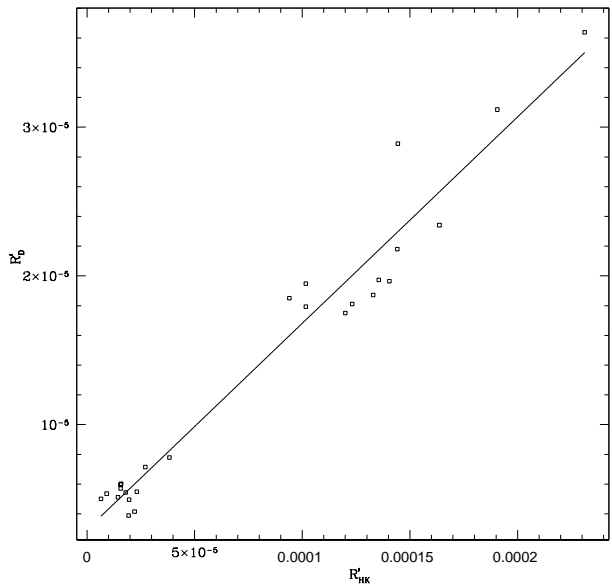}
\caption{$R^\prime_D$ vs $R^\prime_{HK}$ for stars with Balmer lines
  in emission. The solid line is the best linear fit to the data, with
  correlation coefficient $r=0.977$. The excelent correlation
  indicates that $R^\prime_D$ is a good chromospheric activity
  indicator.}
\label{rvsra}
\end{figure}

To study its applicability as an activity indicator, $R^\prime_D$ was
plotted against $R^\prime_{HK}$. The result is presented in
Fig.~\ref{rvsr}, where the filled triangles represent stars which
exhibit the Balmer lines in emission. No correlation seems to be
present when we consider the complete stellar sample. Indeed the
correlation coefficient is $r=0.094$. However, if we consider only the
most active stars --those which exhibit Balmer lines in emission -- an
excelent correlation is found between both indexes. This is shown in
Fig.~\ref{rvsra}, where we plot $R^\prime_D$ vs $R^\prime_{HK}$ only
for these stars. The solid line is the best linear fit to the data:
\begin{equation}
R^\prime_D=2.931\times 10^{-6} + 0.1388 \cdot R^\prime_{HK} \; \; ,
\end{equation}
with correlation coefficient $r=0.978$. This result demonstrates that
$R^\prime_D$ can be used as an activity indicator for active
stars. Note that despite presenting the Balmer lines in emission, not
all stars in Fig.~\ref{rvsra} present the D lines in emission, and
this does not seem to be a condition for the use of $R^\prime_D$ as an
activity indicator. $R^\prime_D$ is particularly useful to study the cooler
stars, which tend to be more active and where the emission in the
region of the H and K Ca II lines is substantially smaller (even an
order of magnitude) than the flux in the region of the D doublet.

\section{Summary}
A program was started in 1999 to study chromospheric activity and
atmospheres of main sequence stars in the Southern Hemisphere. In this
paper we analyze the sodium D lines. We use medium resolution
\emph{echelle} spectra, obtained in CASLEO, which were calibrated in
wavelength and in flux.

We constructed an HR diagram for all the stars in our sample taking
the colour index $(B-V)$ and the visual magnitude $V$ from the
Hipparcos/Tycho catalogues. We found that several stars, in spite of
being classified as dwarfs in the Simbad database, fall outside the
main sequence. We present these stars in Table~\ref{tabla1} and defer
a more detailed analysis for future work. Conversely, in
Table~\ref{tabla2} we show four stars that were not classified as
dwarfs even though they fall within the main sequence.

In Sect.~\ref{pc} we define a pseudo-continuum level useful for all
spectral types. Since in M stars the D lines are formed inside a TiO
molecular band, we consider two windows located quite far from the
doublet.  In each window, the mean value of the ten highest points was
calculated and the pseudo-continuum was defined as the linear
interpolation between both windows. The continuum flux obtained in this way
shows a very tight correlation with colour index $(B-V)$. Since most
observations do no include regions too far from the lines, for stars with
$(B-V)<1.4$ we also defined another pseudo-continuum, which uses a window
located closer to the D lines, and also correlates strongly with $(B-V)$. 
We showed that this correlation can be used to obtain an approximate calibration
in flux in this spectral region.

In Sect.~\ref{sec.an} we describe in detail the computation of the
equivalent width. We use a wavelength interval which range from 14
\AA\ for F6 stars to 40 \AA\ for the coolest stars in our sample. In
M-type stars, the large photospheric wings, blended with deep
molecular bands, difficult an accurate calculation. We find a good 
correlation between  equivalent width and colour index $(B-V)$
for all the range of observations. 

The relation obtained was then used to determine the colour of
dwarf-type stars from the equivalent width of the D lines. We find
values of $(B-V)$ in good agreement with those from the
Hipparcos/Tycho catalogues.  Since equivalent width is a
characteristic of line profiles that do not require high resolution
spectra to be measured, this fact could become a useful tool for
subsequent studies.

Finally, in Sect.~\ref{sec.activ} we study how the flux in the D lines
changes with changing levels of chromospheric activity. We define an
index (N) analogue to Mount Wilson S index and find it is strongly
correlated with colour for stars with $(B-V)<1.1$. This fact produces
an apparent correlation between N and S, which disappears when the
colour dependence is taken into account. On the other hand, in stars
with $(B-V)\geq1.1$, N is independent of colour and is correlated with
the S index. However, the slope between N and S varies when different
stars are studied individually, varying from tight correlations to
cases where no correlation is present. This fact restricts the use of
the N index as an activity indicator when different stars are
compared. However, the N index may be useful when comparing different
activity levels on individual stars, specially in later stars with
little emission in the region of the CaII H \& K lines.

In order to compare activity levels on stars of different spectral
types we define an improved index taking into account the photospheric
contribution to the flux both in the lines and in the continuum
windows. First we construct $R_D$ as the ratio between absolute line
fluxes and stellar luminosity. Then, the photospheric contribution in
the lines bandpasses, which was computed using the basal flux in the D
lines, was substracted from $R_D$ to define $R_D^\prime$. As was
expected, the earlier stars in our sample do not show any sign of
correlation between both indexes. Indeed, it has been known for a long
time that the cores of the Na I D lines in the solar atmosphere remain
dark from center to limb while the Ca II H\&K lines show increasing
line reversal. Only in stars with higher atomic densities, as the
M-dwarfs, the D lines can be expected to have collision rates high
enough to respond to the temperature changes in the chromosphere
\citep[e.g.][]{mdwarves2}. Because of this, earlier theoretical works
\citep[see][]{andretta97,shortdoyle98} have restricted their attention
to this type of stars. However, in this work the new $R_D^\prime$
index was found to correlate well with $R^\prime_{HK}$ for the most
active stars in our sample (those which exhibit the Balmer lines in
emission), even though some of these stars do not present a line
reversal at the core of the D lines. Therefore, $R^\prime_D$ is also
a good activity indicator for these stars. For the rest of the
spectra, no correlation was found.

The CCD and data acquisition system at CASLEO has been partly financed
by R. M. Rich through U.S. NSF grant AST-90-15827. This research has
made use of the SIMBAD database, operated at CDS, Strasbourg,
France. We thankfully acknowledge the comments and suggestions of the
referee (Ian Short), which helped us to improve our original
manuscript.

\clearpage
\renewcommand{\arraystretch}{1}
\pagestyle{empty}

\setcounter{table}{3}
\begin{landscape}
\begin{table}
\caption{\label{todas}List of stars used in the analysis and average
  measurements. Nobs is the number of times the star has been
  observed. $V$ is the visual magnitude in the UBV system. [Fe/H] is
  the iron abundance relative to the Sun, obtained from references
  (a): Cayrel de Strobel et al. (2001); (b): Nordstr\"om et
  al. (2004); (c): Cayrel de Strobel et al. (1997). $<F_{cont}>$ is
  the continuum flux (see Sect. 3.1). $<\bar{F}>$ is the mean flux in
  a window 20 \AA\ wide around 5840 \AA. $<W_\lambda>$ is the
  equivalent width of the sodium doublet and $<W_{approx}>$ is the
  equivalent width computed using $\bar{F}$ as pseudo-continuum (see
  Sect. 4.1), in angstroms. Cols.~10 and 11 give the average absolute
  flux in a 1 \AA\ window centered at the D1 and D2 lines,
  respectively. All fluxes are in units of
  [$\mathrm{erg~cm^{-2}~s^{-1}}$].  $<\mathrm{N}>$ is the value of the
  N index, calculated as explained in Sect. 5.1. Columns 13 and 14
  give average value for the $R_D$ index and the corresponding
  photospheric correction $R^{phot}_D$ (see Sect. 5.2). Finally, in
  the last column we present the mean values of
  $R^\prime_{HK}$. Colour index $(B-V)$, magnitude $V$ and spectral
  type have been obtained from the Hipparcos/Tycho catalogues
  (Perryman et al. 1997; Hoeg et al. 1997).}  {\footnotesize
\begin{tabular}{l c c c c c c c c c c c c c c c}
\hline
Star&Nobs&Sp.
type&$(B-V)$&$V$&[Fe/H]&$<F_{cont}>$&$<\bar{F}>$&$<W_\lambda>$&$<W_{approx}>$&$<F_{D1}>$&$<F_{D2}>$&$<\mathrm{N}>$&$<R_D>$&$R_D^{phot}$&$<R^\prime_{HK}>$\\
\hline
hd28246&5&F6V&0.457&6.380&0.02$^b$&9.651e-12&9.591e-12&1.189&1.108&6.330e-12&5.7
3 9 e - 1 2 & 1.251&1.825e-04&1.761e-04&5.737e-05\\ 
hd38393&12&F7V&0.481&3.590&-0.12$^a$&1.293e-10&1.286e-10&1.222&1.153&8.595e-11&7
. 8 6 2e-11&1.272&1.897e-04&1.780e-04&3.084e-05\\ 
hd120136&3&F6IV&0.508&4.500&0.32$^a$&5.665e-11&5.618e-11&1.602&1.500&3.622e-11&3
.270e-11&1.217&1.829e-04&1.037e-04&4.078e-05\\ 
hd16673&9&F6V&0.524&5.790&-0.01$^a$&1.711e-11&1.704e-11&1.243&1.192&1.098e-11&9.
567e-12&1.201&1.784e-04&1.529e-04&4.929e-05\\ 
hd114762&3&F9V&0.525&7.300&-0.82$^a$&4.346e-12&4.325e-12&1.069&1.008&2.913e-12&2
.641e-12&1.278&1.938e-04&1.716e-04&2.693e-05\\ 
hd35850&5&F7V&0.553&6.300&0.00$^a$&1.063e-11&1.057e-11&1.374&1.296&7.466e-12&6.6
73e-12&1.330&1.940e-04&1.198e-04&1.376e-04\\ 
hd17051&5&G0V&0.561&5.400&-0.04$^a$&2.536e-11&2.505e-11&1.679&1.526&1.423e-11&1.
233e-11&1.047&1.582e-04&1.176e-04&4.635e-05\\ 
hd45067&11&F8V&0.564&5.880&-0.16$^a$&1.572e-11&1.562e-11&1.340&1.255&9.635e-12&8
.603e-12&1.160&1.686e-04&4.831e-05&2.549e-05\\ 
hd52265&2&G0III-IV&0.572&6.290&0.21$^a$&1.108e-11&1.091e-11&1.541&1.347&5.911e-1
2&5.201e-12&1.003&1.490e-04&9.101e-05&2.567e-05\\ 
hd19994&9&F8V&0.575&5.070&0.15$^a$&3.438e-11&3.385e-11&1.736&1.543&1.932e-11&1.7
17e-11&1.062&1.587e-04&4.529e-05&3.132e-05\\ 
hd130948&6&G1V&0.576&5.860&-0.20$^a$&1.640e-11&1.622e-11&1.885&1.744&9.382e-12&8
.210e-12&1.072&1.582e-04&1.445e-04&5.599e-05\\ 
hd75289&8&G0Ia&0.578&6.350&0.28$^a$&1.054e-11&1.039e-11&1.651&1.472&6.067e-12&5.
247e-12&1.074&1.596e-04&8.565e-05&3.365e-05\\ 
hd215768&5&G0V&0.589&7.490&-0.20$^b$&3.678e-12&3.642e-12&1.466&1.342&2.083e-12&1
. 7 9 5 e - 1 2 &1.054&1.550e-04&1.193e-04&6.666e-05\\ 
hd213240&2&G0/1V&0.603&6.810&0.13$^b$&6.975e-12&6.907e-12&1.975&1.856&3.737e-12&
3 . 2 1 1 e - 12&0.996&1.484e-04&5.347e-05&2.894e-05\\ 
hd43587&7&F9V&0.610&5.700&-0.08$^a$&1.932e-11&1.908e-11&1.812&1.656&1.103e-11&9.
575e-12&1.067&1.583e-04&8.101e-05&2.246e-05\\ 
hd197076&6&G5V&0.611&6.430&-0.20$^b$&9.906e-12&9.812e-12&1.560&1.440&5.488e-12&4
. 6 9 6 e - 1 2 &1.028&1.533e-04&1.334e-04&2.160e-05\\ 
hd202996&2&G0V&0.614&7.460&0.00$^b$&3.735e-12&3.690e-12&1.778&1.629&2.355e-12&2.
0 4 7 e - 1 2&1.179&1.711e-04&5.219e-05&6.028e-05\\ 
hd45270&6&G1V&0.614&6.530&-0.14$^b$&9.022e-12&8.913e-12&1.770&1.620&5.030e-12&4.
4 6 4 e - 1 2 &1.052&1.567e-04&1.141e-04&8.372e-05\\ 
hd165185&10&G5V&0.615&5.940&-0.06$^a$&1.540e-11&1.523e-11&1.727&1.589&8.778e-12&
7.63 2e-12&1.066&1.573e-04&1.203e-04&6.079e-05\\ 
hd48189&6&G1.5V&0.624&6.150&-0.18$^b$&1.259e-11&1.248e-11&1.878&1.770&7.036e-12&
6 . 3 7 0 e - 1 2&1.065&1.559e-04&8.756e-05&8.837e-05\\ 
hd147513&11&G5V&0.625&5.370&0.03$^a$&2.656e-11&2.618e-11&1.754&1.577&1.508e-11&1
.271e-11&1.046&1.576e-04&1.201e-04&6.066e-05\\ 
hd150433&7&G0&0.631&7.210&-0.47$^b$&4.702e-12&4.649e-12&1.753&1.613&2.666e-12&2.
294 e - 1 2 & 1.055&1.530e-04&1.185e-04&2.568e-05\\ 
hd219709&7&G2/3V&0.632&7.500&-0.02$^a$&3.607e-12&3.572e-12&1.925&1.808&1.922e-12
&1.6 48e-12&0.990&1.438e-04&8.602e-05&4.212e-05\\ 
hd30495&11&G3V&0.632&5.490&-0.13$^a$&2.326e-11&2.304e-11&1.696&1.576&1.228e-11&1
.047e-11&0.978&1.439e-04&1.187e-04&6.058e-05\\ 
hd202628&9&G2Va&0.637&6.750&-0.14$^a$&7.346e-12&7.261e-12&1.954&1.812&3.955e-12&
3.401e-12&1.001&1.480e-04&1.141e-04&4.981e-05\\ 
hd38858&5&G4V&0.639&5.970&-0.25$^b$&1.511e-11&1.492e-11&1.872&1.710&8.241e-12&7.
0 4 6 e - 1 2 & 1.011&1.498e-04&1.279e-04&2.778e-05\\ 
hd20766&12&G2.5V&0.641&5.530&-0.22$^a$&2.242e-11&2.225e-11&1.833&1.742&1.199e-11
&9.930e-12&0.978&1.430e-04&1.384e-04&4.897e-05\\ 
hd59967&5&G4V&0.641&6.660&-0.19$^b$&8.012e-12&7.884e-12&1.826&1.629&4.420e-12&3.
7 6 3 e - 1 2 & 1.021&1.512e-04&1.214e-04&7.739e-05\\ 
hd187923&10&G0V&0.642&6.160&-0.20$^a$&1.368e-11&1.350e-11&1.809&1.646&7.548e-12&
6.660e-12&1.038&1.655e-04&4.708e-05&2.116e-05\\ 
hd19467&10&G3V&0.645&6.970&-0.13$^b$&5.923e-12&5.861e-12&1.736&1.607&3.277e-12&2
. 8 7 4 e - 1 2 &1.039&1.508e-04&7.483e-05&2.271e-05\\ 
hd189567&8&G3V&0.648&6.070&-0.30$^a$&1.395e-11&1.377e-11&1.752&1.595&7.994e-12&6
.906e-12&1.068&1.592e-04&1.008e-04&3.051e-05\\ 
hd210918&11&G5V&0.648&6.230&-0.18$^a$&1.154e-11&1.139e-11&1.788&1.623&6.353e-12&
5.686e-12&1.043&1.491e-04&7.484e-05&2.604e-05\\ 
hd173560&9&G3/5V&0.649&8.690&-0.21$^b$&1.272e-12&1.248e-12&1.667&1.427&7.344e-13
& 6 . 3 8 5 e - 13&1.079&1.637e-04&4.757e-05&3.793e-05\\ 
hr6060&15&G2Va&0.652&5.490&0.05$^a$&2.369e-11&2.331e-11&1.962&1.765&1.221e-11&1.
033e-11&0.951&1.408e-04&9.134e-05&2.526e-05\\ 
hd11131&9&G0&0.654&6.720&-0.06$^a$&7.558e-12&7.474e-12&1.726&1.587&4.098e-12&3.4
39e-12&0.997&1.460e-04&1.038e-04&7.184e-05\\ 
hd20619&11&G1.5V&0.655&7.050&-0.20$^a$&5.339e-12&5.293e-12&1.738&1.632&2.916e-12
&2.486e-12&1.012&1.417e-04&1.212e-04&3.218e-05\\ 
hd217343&2&G3V&0.655&7.470&-0.18$^b$&3.768e-12&3.711e-12&1.995&1.810&2.075e-12&1
. 8 0 5 e - 1 2 &1.030&1.499e-04&1.060e-04&1.057e-04\\ 
hd4308&12&G5V&0.655&6.550&-0.47$^a$&8.907e-12&8.808e-12&1.652&1.513&4.651e-12&3.
991e-12&0.970&1.431e-04&9.757e-05&2.719e-05\\ 
hd1835&13&G3V&0.659&6.390&-0.01$^a$&1.067e-11&1.049e-11&2.116&1.921&5.225e-12&4.
551e-12&0.917&1.393e-04&9.379e-05&6.410e-05\\ 
hd213941&9&G5V&0.670&7.580&-0.42$^a$&3.335e-12&3.313e-12&2.057&1.977&1.759e-12&1
.508e-12&0.979&1.386e-04&1.031e-04&3.985e-05\\ 

\end{tabular}
}
\end{table}
\clearpage
\begin{table}
\contcaption{}
{\footnotesize
\begin{tabular}{l c c c c c c c c c c c c c c c}
\hline
Star&Nobs&Sp.
type&$(B-V)$&$V$&[Fe/H]&$<F_{cont}>$&$<\bar{F}>$&$<W_\lambda>$&$<W_{approx}>$&$<F_{D1}>$&$<F_{D2}>$&$<\mathrm{N}>$&$<R_D>$&$R_D^{phot}$&$<R^\prime_{HK}>$\\
\hline
hd197214&10&G5V&0.671&6.950&-0.5$^b$&6.138e-12&6.048e-12&1.983&1.804&3.224e-12&2
. 7 1 9 e - 1 2&0.968&1.411e-04&1.186e-04&3.608e-05\\ 
hd172051&13&G5V&0.673&5.850&-0.29$^b$&1.699e-11&1.677e-11&2.078&1.923&8.668e-12&
7 . 3 6 7 e - 1 2&0.944&1.380e-04&1.266e-04&2.725e-05\\ 
hd19034&8&G5&0.677&8.080&-0.39$^b$&2.167e-12&2.148e-12&1.870&1.767&1.128e-12&9.4
4 3 e - 1 3 & 0 .957&1.389e-04&1.297e-04&2.861e-05\\ 
hd203019&3&G5V&0.687&7.840&0.07$^b$&2.710e-12&2.652e-12&2.330&2.078&1.256e-12&1.
1 2 3 e - 1 2 &0.878&1.272e-04&9.105e-05&9.492e-05\\ 
hd202917&5&G5V&0.690&8.650&-0.16$^b$&1.269e-12&1.250e-12&2.377&2.198&6.377e-13&5
. 5 0 5 e - 1 3 &0.936&1.338e-04&1.183e-04&1.416e-04\\ 
hd12759&4&G3V&0.694&7.300&-0.13$^b$&4.585e-12&4.501e-12&2.245&2.026&2.427e-12&2.
0 9 3 e - 1 2 &0.986&1.465e-04&2.882e-05&7.779e-05\\ 
hd128620&7&G2V&0.710&-0.010&0.22$^a$&3.844e-09&3.777e-09&2.123&1.912&1.943e-09&1
. 5 9 1e-09&0.919&1.355e-04&4.082e-05&2.221e-05\\ 
hd41824&11&G6V&0.712&6.600&-0.09$^b$&9.098e-12&8.924e-12&2.456&2.231&4.401e-12&3
.864 e - 1 2 &0.909&1.394e-04&3.642e-05&1.074e-04\\ 
hd117176&3&G4V&0.714&4.970&-0.11$^a$&3.835e-11&3.783e-11&1.947&1.780&2.074e-11&1
.897e-11&1.036&1.492e-04&2.157e-05&1.581e-05\\ 
hd3443&9&K1V&0.715&5.570&-0.16$^c$&2.212e-11&2.186e-11&2.440&2.305&1.051e-11&8.7
90e-12&0.872&1.259e-04&5.059e-05&2.837e-05\\ 
hd3795&9&G3/5V&0.718&6.140&-0.70$^a$&1.299e-11&1.288e-11&1.692&1.584&7.208e-12&6
.159e-12&1.029&1.472e-04&2.478e-05&2.282e-05\\ 
hd203244&1&G5V&0.723&6.980&-0.21$^a$&6.150e-12&6.069e-12&2.284&2.128&3.265e-12&2
.754e-12&0.979&1.434e-04&1.010e-04&7.445e-05\\ 
hd10700&11&G8V&0.727&3.490&-0.59$^a$&1.490e-10&1.475e-10&2.344&2.227&6.720e-11&5
.630e-11&0.829&1.180e-04&1.243e-04&3.117e-05\\ 
hd152391&11&G8V&0.749&6.650&-0.06$^b$&8.579e-12&8.407e-12&2.613&2.379&3.882e-12&
3.26 3 e - 1 2&0.833&1.238e-04&9.103e-05&6.268e-05\\ 
hd36435&5&G6/8V&0.755&6.990&-0.02$^a$&6.071e-12&5.984e-12&2.425&2.255&2.798e-12&
2.433e-12&0.862&1.235e-04&8.948e-05&6.689e-05\\ 
hd13445&5&K1V&0.812&6.120&-0.21$^a$&1.365e-11&1.338e-11&3.400&3.147&4.822e-12&3.
947e-12&0.643&8.931e-05&8.795e-05&3.455e-05\\ 
hd26965&10&K1V&0.820&4.430&-0.25$^a$&6.876e-11&6.776e-11&3.413&3.228&2.547e-11&2
.081e-11&0.673&9.885e-05&8.235e-05&2.275e-05\\ 
hd177996&9&K1V&0.862&7.890&-0.28$^b$&2.876e-12&2.798e-12&4.034&3.699&1.131e-12&9
.836 e - 1 3 &0.735&1.048e-04&3.767e-05&7.872e-05\\ 
hd17925&7&K1V&0.862&6.050&0.10$^a$&1.648e-11&1.608e-11&3.898&3.596&5.892e-12&4.9
04e-12&0.655&9.825e-05&6.478e-05&8.305e-05\\ 
hd22049&13&K2V&0.881&3.720&-0.14$^a$&1.318e-10&1.292e-10&3.905&3.653&4.355e-11&3
.628e-11&0.606&8.335e-05&6.921e-05&5.299e-05\\ 
hd128621&8&K1V&0.900&1.350&0.24$^a$&1.220e-09&1.183e-09&4.576&4.216&3.709e-10&3.
049e-10&0.554&7.802e-05&3.906e-05&2.154e-05\\ 
gl349&12&K3V&1.002&7.200&-0.15$^b$&5.535e-12&5.373e-12&6.086&5.667&1.353e-12&1.1
42e- 1 2 & 0 .451&5.558e-05&4.579e-05&5.335e-05\\ 
gl542&9&K3V&1.017&6.660&0.26$^a$&9.770e-12&9.361e-12&7.693&7.154&1.959e-12&1.678
e-12&0.372&4.834e-05&2.846e-05&1.566e-05\\ 
hd188088&6&K3/4V&1.017&6.220&--&1.499e-11&1.433e-11&7.920&7.369&3.456e-12&2.903e-12&0.425&5.636e-05&1.315e-05&4.730e-05\\ 
hd131977&9&K4V&1.024&5.720&0.03&2.381e-11&2.322e-11&8.199&7.903&4.477e-12&3.802e-12&0.348&4.588e-05&4.552e-05&2.660e-05\\ 
gl610&9&K3/4V&1.043&7.390&-0.03$^a$&4.892e-12&4.719e-12&7.510&7.054&1.128e-12&9.
638e-13&0.428&5.266e-05&1.849e-05&1.505e-05\\ 
hd32147&10&K3V&1.049&6.220&0.34$^a$&1.468e-11&1.415e-11&7.691&7.232&2.773e-12&2.
353e-12&0.349&4.360e-05&2.653e-05&1.978e-05\\ 
gl845&8&K4.5V&1.056&4.690&-0.23$^a$&6.041e-11&5.768e-11&7.782&7.203&1.179e-11&1.
008e-11&0.362&4.505e-05&3.618e-05&2.748e-05\\ 
gl435&8&K4/5V&1.064&7.770&--&3.576e-12&3.467e-12&7.456&7.062&7.230e-13&6.108e-13&0.373&4.640e-05&4.852e-05&2.499e-05\\ 
hd216803&2&K4V&1.094&6.480&--&1.181e-11&1.159e-11&6.814&6.562&2.636e-12&2.258e-12&0.414&4.993e-05&3.080e-05&6.597e-05\\ 
hd156026&7&K5V&1.144&6.330&-0.21$^a$&1.398e-11&1.357e-11&8.456&8.105&2.623e-12&2
.313e-12&0.353&4.114e-05&2.798e-05&3.156e-05\\ 
gl416&7&K4V&1.203&9.060&--&1.178e-12&1.121e-12&10.178&9.482&2.018e-13&1.851e-13&0.329&3.623e-05&1.301e-05&2.745e-05\\ 
gl517&7&K5Ve&1.210&9.240&-0.15$^c$&9.272e-13&8.849e-13&9.795&9.116&2.115e-13&1.9
72e-13&0.441&4.464e-05&1.877e-05&1.643e-04\\ 
hd157881&9&K5&1.359&7.540&-0.20$^a$&5.122e-12&4.868e-12&13.782&13.040&8.167e-13&
7.218e-13&0.300&2.634e-05&3.709e-06&3.227e-05\\ 
gl526&5&M1.5&1.435&8.460&--&2.297e-12&--&15.084&--&3.925e-13&3.433e-13&0.320&2.379e-05&4.811e-06&1.880e-05\\ 
gl536&5&M1&1.461&9.710&--&7.087e-13&--&14.472&--&1.436e-13&1.307e-13&0.387&2.547e-05&2.607e-06&1.920e-05\\ 
gl1&9&M1.5&1.462&8.560&--&2.062e-12&--&15.579&--&2.574e-13&2.163e-13&0.230&1.520e-05&4.825e-06&6.587e-06\\ 
hd180617&8&M2.5&1.464&9.120&--&1.220e-12&--&16.324&--&1.837e-13&1.661e-13&0.287&1.866e-05&4.279e-06&2.185e-05\\ 
gl479&6&M2&1.470&10.650&--&3.116e-13&--&16.527&--&5.484e-14&5.032e-14&0.338&2.246e-05&5.738e-06&3.258e-05\\ 
hd36395&8&M1.5&1.474&7.970&0.60$^c$&3.719e-12&--&15.876&--&6.313e-13&5.648e-13&0
.32 2&2.133e-05&1.291e-06&4.688e-05\\ 
hd42581&8&M1/2V&1.487&8.150&--&3.000e-12&--&14.864&--&5.104e-13&4.495e-13&0.320&
1.895e-05&1.113e-06&2.533e-05\\ 
gl388&6&M3.5Ve&1.540&9.430&--&9.761e-13&--&22.636&--&2.003e-13&2.289e-13&0.440&1.987e-05&1.501e-06&1.122e-04\\ 
gl699&11&M4Ve&1.570&9.540&--&9.174e-13&--&26.649&--&7.240e-14&6.531e-14&0.150&4.653e-06&4.012e-06&1.006e-05\\ 
gl551&13&M5.5Ve&1.807&11.010&--&2.303e-13&--&28.131&--&4.582e-14&5.835e-14&0.452&5.566e-06&2.657e-08&1.883e-05\\ 

\end{tabular}
}
\end{table}
\end{landscape}


\begin{thebibliography}{}

\bibitem[\protect\citeauthoryear{{Allen}}{{Allen}}{1964}]{allenviejo}
{Allen} C.~W.,  1964, {Astrophysical Quantities}.
Astrophysical Quantities, London: Athlone Press (2nd edition), 1964

\bibitem[\protect\citeauthoryear{{Andretta}, {Doyle} \& {Byrne}}{{Andretta}
  et~al.}{1997}]{andretta97}
{Andretta} V.,  {Doyle} J.~G.,    {Byrne} P.~B.,  1997, \aap, 322, 266

\bibitem[\protect\citeauthoryear{Buccino \& Mauas}{Buccino \&
  Mauas}{2007}]{magnesio}
Buccino A.,  Mauas P.~J.~D.,  2007, {\emph{in preparation}}

\bibitem[\protect\citeauthoryear{{Cayrel de Strobel}, {Soubiran}, {Friel},
  {Ralite} \& {Francois}}{{Cayrel de Strobel} et~al.}{1997}]{cayrel97}
{Cayrel de Strobel} G.,  {Soubiran} C.,  {Friel} E.~D.,  {Ralite} N.,
  {Francois} P.,  1997, \aaps, 124, 299

\bibitem[\protect\citeauthoryear{{Cayrel de Strobel}, {Soubiran} \&
  {Ralite}}{{Cayrel de Strobel} et~al.}{2001}]{cayrel2001}
{Cayrel de Strobel} G.,  {Soubiran} C.,    {Ralite} N.,  2001, \aap, 373, 159

\bibitem[\protect\citeauthoryear{Cincunegui, D\'iaz \& Mauas}{Cincunegui
  et~al.}{2006}]{indices}
Cincunegui C.,  D\'iaz R.~F.,    Mauas P.~J.~D.,  2007, \aap,
  {\emph{in press}}

\bibitem[\protect\citeauthoryear{{Cincunegui}, {D{\'{\i}}az} \&
  {Mauas}}{{Cincunegui} et~al.}{2007}]{proxima}
{Cincunegui} C.,  {D{\'{\i}}az} R.~F.,    {Mauas} P.~J.~D.,  2007, \aap, 461,
  1107

\bibitem[\protect\citeauthoryear{Cincunegui \& Mauas}{Cincunegui \&
  Mauas}{2002}]{cycles}
Cincunegui C.,  Mauas P.~J.~D.,  2002, in ESA SP-477: Solspa 2001, Proceedings
  of the Second Solar Cycle and Space Weather Euroconference {Cycles in other
  stars}.
pp 91--94

\bibitem[\protect\citeauthoryear{Cincunegui \& Mauas}{Cincunegui \&
  Mauas}{2004}]{library}
Cincunegui C.,  Mauas P.~J.~D.,  2004, \aap, 414, 699

\bibitem[\protect\citeauthoryear{{Henry}, {Soderblom}, {Donahue} \&
  {Baliunas}}{{Henry} et~al.}{1996}]{CTIO}
{Henry} T.~J.,  {Soderblom} D.~R.,  {Donahue} R.~A.,    {Baliunas} S.~L.,
  1996, \aj, 111, 439

\bibitem[\protect\citeauthoryear{{Hoeg}, {B{\"a}ssgen}, {Bastian}, {Egret},
  {Fabricius}, {Gro{\ss}mann}, {Halbwachs}, {Makarov}, {Perryman},
  {Schwekendiek}, {Wagner} \& {Wicenec}}{{Hoeg} et~al.}{1997}]{tycho}
{Hoeg} E.,  {B{\"a}ssgen} G.,  {Bastian} U.,  {Egret} D.,  {Fabricius} C.,
  {Gro{\ss}mann} V.,  {Halbwachs} J.~L.,  {Makarov} V.~V.,  {Perryman}
  M.~A.~C.,  {Schwekendiek} P.,  {Wagner} K.,    {Wicenec} A.,  1997, \aap,
  323, L57

\bibitem[\protect\citeauthoryear{{Johnson}}{{Johnson}}{1966}]{johnson66}
{Johnson} H.~L.,  1966, \araa, 4, 193

\bibitem[\protect\citeauthoryear{{Mauas}}{{Mauas}}{2000}]{mdwarves2}
{Mauas} P.~J.~D.,  2000, \apj, 539, 858

\bibitem[\protect\citeauthoryear{{Mauas} \& {Falchi}}{{Mauas} \&
  {Falchi}}{1994}]{adleo1}
{Mauas} P.~J.~D.,  {Falchi} A.,  1994, \aap, 281, 129

\bibitem[\protect\citeauthoryear{{Nordstr{\"o}m}, {Mayor}, {Andersen},
  {Holmberg}, {Pont}, {J{\o}rgensen}, {Olsen}, {Udry} \&
  {Mowlavi}}{{Nordstr{\"o}m} et~al.}{2004}]{nordstrom2004}
{Nordstr{\"o}m} B.,  {Mayor} M.,  {Andersen} J.,  {Holmberg} J.,  {Pont} F.,
  {J{\o}rgensen} B.~R.,  {Olsen} E.~H.,  {Udry} S.,    {Mowlavi} N.,  2004,
  \aap, 418, 989

\bibitem[\protect\citeauthoryear{{Noyes}, {Hartmann}, {Baliunas}, {Duncan} \&
  {Vaughan}}{{Noyes} et~al.}{1984}]{noyes84}
{Noyes} R.~W.,  {Hartmann} L.~W.,  {Baliunas} S.~L.,  {Duncan} D.~K.,
  {Vaughan} A.~H.,  1984, \apj, 279, 763

\bibitem[\protect\citeauthoryear{Perryman et al.}%
  {1997}]{hipparcos}{Perryman} M.~A.~C., {Lindegren} L., {Kovalevsky}
  J., {Hoeg} E., {Bastian} U., {Bernacca} P.~L., {Cr{\'e}z{\'e}} M.,
  {Donati} F., {Grenon} M., {van Leeuwen} F., {van der Marel} H.,
  {Mignard} F., {Murray} C.~A., {Le Poole} R.~S., {Schrijver} H.,
  {Turon} C., {Arenou} F., {Froeschl{\'e}} M., {Petersen} C.~S., 1997,
  \aap, 323, L49

\bibitem[\protect\citeauthoryear{{Short} \& {Doyle}}{{Short} \&
  {Doyle}}{1998}]{shortdoyle98}
{Short} C.~I.,  {Doyle} J.~G.,  1998, \aap, 336, 613

\bibitem[\protect\citeauthoryear{{Tripicchio}, {Severino}, {Covino},
  {Terranegra} \& {Garcia Lopez}}{{Tripicchio} et~al.}{1997}]{tripi}
{Tripicchio} A.,  {Severino} G.,  {Covino} E.,  {Terranegra} L.,    {Garcia
  Lopez} R.~J.,  1997, \aap, 327, 681

\bibitem[\protect\citeauthoryear{{Vaughan}, {Preston} \& {Wilson}}{{Vaughan}
  et~al.}{1978}]{MW}
{Vaughan} A.~H.,  {Preston} G.~W.,    {Wilson} O.~C.,  1978, \pasp, 90, 267

\end{thebibliography}
\end{document}